\documentclass[aps, prl, twocolumn, 
amssymb,superscriptaddress]{revtex4}
\usepackage{graphicx, bm, amsmath, amsfonts,amssymb}
\usepackage{subfigure}

\def\t#1{\widetilde{#1}}

\begin{document}

\title{A universal form for quark and lepton mass matrices}

\author{Zheng-Cheng Gu}
\affiliation{Perimeter Institute for Theoretical Physics, Waterloo, Ontario, N2L2Y5, Canada}
\author{John Preskill}
\affiliation{Institute for Quantum Information and Matter, Caltech, Pasadena, CA 91125, USA}

\begin{abstract}
We propose a universal form for quark and lepton mass matrices, which applies in a ``leading order'' approximation where $CP$-violating phases are ignored. Down-quark mass ratios are successfully predicted in our scheme using the measured CKM mixing angles as input. Assuming an additional discrete symmetry in the neutrino sector, we obtain the ``golden ratio'' pattern in the leading-order PMNS mixing matrix; in addition we predict an inverted neutrino mass hierarchy with $m_1\simeq m_2 \simeq74 ~meV$, $m_3\simeq 55~ meV$, and neutrinoless double beta decay mass parameter $m_{0\nu\beta\beta}\simeq 33~ meV$. We also predict that the $CP$-violating angle in the neutrino sector is close to the maximal value $\delta=\pm\pi/2$, and that the diagonal phases in the PMNS matrix are $\alpha_1\simeq 0$, $\alpha_2\simeq\pi$.
\end{abstract}

\maketitle

\emph{Introduction.---} The masses and mixing angles of quarks \cite{CKM1,CKM2,CKM3,CKM5} and leptons \cite{neutrinoexp1,neutrinoexp2,neutrinoexp3,neutrinoexp4,neutrinoexp5,neutrinoexp6,neutrinoexp7,neutrinoexp8,neutrinoexp9,neutrinoexp10,neutrinoexp11,neutrinoexp12}, fundamental parameters of our universe, are utterly mysterious. Many attempts have been made to explain or relate the quark and lepton mass matrices, for example by invoking quark-neutrino complementarity \cite{complementarity1,complementarity2} or (discrete) flavor symmetry \cite{symmetry1,symmetry2,symmetry3}, but with limited success. In the discrete flavor symmetry approach in particular, symmetries enforce constraints on neutrino mixing angles which are in reasonable agreement with experimental observations \cite{masssymmetry3,masssymmetry4}, but neutrino mass ratios are not constrained, and these symmetries are not respected in the quark sector.

In this paper, we propose an Ansatz for quark and lepton mass matrices that accounts well for all observed quark and lepton masses and mixing angles in terms of a small number of free parameters, and also makes predictions which can be tested in future neutrino oscillation experiments. We assume that in each of four sectors (up quarks, down quarks, charged leptons, and (very heavy) right-handed neutrinos), the mass matrix has the same universal form. For quarks, ignoring $CP$-violating phases and overall mass scales, this universal matrix has two free parameters in each of the up and down sectors. These parameters are fixed by observed mass ratios, so that all Cabibbo-Kobayashi-Maskawa (CKM) mixing angles are predicted. $CP$ violation can also be accommodated, and the predicted CKM matrix is actually quite insensitive to the mass ratios in the up sector. For leptons, ignoring $CP$-violating phases, there are two free parameters in the charged sector, again fixed by observed mass ratios, but no free parameters in the neutrino sector, so that all neutrino masses and mixing angles are predicted. The $CP$ violation in the lepton sector can also be predicted up to a sign ambiguity by assuming an unbroken $\mu-\tau$ $\mathbb{Z}_2$ antiunitary symmetry, and the predicted Pontecorvo-Maki-Nakagawa-Sakata (PMNS) mixing matrix is insensitive to the charged-lepton mass ratios.

Aside from its predictive power, our approach is appealing because it provides a unified description of the quark and lepton sectors. Our Ansatz was inspired by recent speculations regarding the origin of the three generations of neutrinos and their mass mixing matrix \cite{Guneutrino}, but this paper is logically independent of that earlier work. Our Ansatz is purely phenomenological; for now we offer no deeper justification for our assumptions.

\emph{Universal mass matrix.---} In the quark sector, our predicted CKM mixing angles are not very sensitive to the values of $CP$-violating phases or to the form of the up-quark mass matrix. Therefore, we will begin by considering a ``leading order'' (LO) approximation in which the unitary matrix that diagonalizes the up-quark mass matrix is assumed to be the identity transformation, and in which the down-quark mass matrix is assumed to be real. Later, we will discuss how the predicted CKM matrix is ``corrected'' when $CP$-violating phases are included and the diagonalization of the up-quark mass matrix is treated properly.

In the LO approximation we propose that the down-quark mass matrix (up to an overall mass scale) has the form
\begin{eqnarray}
m(\lambda,\lambda^\prime)=\left(
    \begin{array}{ccc}
      1 &  -2 &  -\sqrt{2} \\
      -2 & \lambda &  -\lambda\sqrt{2} \\
       -\sqrt{2} &  -\lambda\sqrt{2} & \lambda^\prime \\
    \end{array}
  \right),\label{mass-matrix}
\end{eqnarray}
where the two adjustable parameters $\lambda>0$ and $\lambda^\prime>0$ are determined by the down-quark mass ratios. In general, the down-quark mass matrix need not be real or Hermitian, and is diagonalized by applying different unitary transformations acting on the left and right, {\em i.e.}, can be expressed as $V_L {\rm{diag}}(m_d,m_s,m_b) V^\dagger_R$. In our Ansatz, though,  $m(\lambda,\lambda^\prime)$ is real and symmetric so that $V_L=V_R=V$. When $CP$ violation is included, we continue to assume the left-right symmetry $V_L=V_R\equiv V$ in the down-quark sector, thus $m_{CP}(\lambda,\lambda^\prime)$ is a Hermitian matrix \cite{NeilCKM1,NeilCKM2}. We assume that the up-quark mass matrix also has the form Eq.(\ref{mass-matrix}), with the values of $\lambda,\lambda'$ determined by up-quark mass ratios. Since these up-quark mass ratios are large compared to the corresponding down-quark mass ratios, CKM mixing angles are not much affected by the up-quark corrections.

Likewise, in the lepton sector we will first consider an LO approximation in which the mass matrix for heavy right-handed neutrinos is real and the transformation diagonalizing the charged-lepton mass matrix is trivial; later we discuss how the PMNS mixing matrix is corrected by $CP$-violating phases and proper diagonalization of the charged-lepton mass matrix. In the LO approximation we propose that the right-handed neutrino mass matrix (up to an overall mass scale) has the form Eq.(\ref{mass-matrix}), but where now $\lambda = \lambda' = 1$, so that the mass matrix has an enhanced symmetry which we will discuss below. This is a Majorana mass matrix, which must be symmetric, and can be expressed as $U {\rm{diag}}(M_1,M_2,M_3) U^T$, where $U$ is unitary and the eigenvalues are the right-handed neutrino masses. The light left-handed neutrinos acquire mass via the seesaw mechanism, and to enhance predictive power we assume that the off-diagonal Dirac mass matrix coupling heavy and light neutrinos is maximally symmetric, {\em i.e.}, proportional to the identity matrix. We assume that the charged-lepton mass matrix also has the form Eq.(\ref{mass-matrix}), with the values of $\lambda,\lambda'$ determined by charged-lepton mass ratios. Since these mass ratios are large compared to the corresponding neutrino mass ratios, PMNS mixing angles are not much affected by these charged-lepton corrections.

\emph{LO approximation for CKM matrix.---}
To find the CKM quark mixing matrix in the LO approximation, we may express $m(\lambda,\lambda^\prime)$ as $V {\rm{diag}}(m_d,m_s,m_b) V^\dagger$, and fix the values of $\lambda$ and $\lambda'$ using the experimentally observed mass ratios $m_s/m_d$ and $m_b/m_s$. But since the observed quark mass ratios have larger uncertainties than the CKM matrix itself, it may be preferable to use the opposite strategy --- fitting $\lambda,\lambda^\prime$ to the CKM matrix, thereby predicting the quark mass ratios (defined at the electroweak symmetry-breaking energy scale).

Choosing $\lambda=10$ and $\lambda^\prime=350$, we find the absolute values of the entries in the CKM matrix
\begin{eqnarray}
|V|=
\left(
                     \begin{array}{ccc}
                        |V_{ud}| & |V_{us}|  & |V_{ub}|\\
                        |V_{cd}| & |V_{cs}|  & |V_{cb}| \\
                        |V_{td}| & |V_{ts}|  & |V_{tb}| \\
                      \end{array}
                    \right)
\simeq
                    \left(
                     \begin{array}{ccc}
                        .974 & .225  & .004\\
                        .225 & .973 &  .041 \\
                        .013 & .040 &  .999 \\
                      \end{array}
                    \right),\label{mixing}
\end{eqnarray}
which are reasonably close to the experimental values \cite{CKM5}:
\begin{eqnarray}
|V_{{\rm{CKM}}}|\simeq
                    \left(
                     \begin{array}{ccc}
                        .974 & .225  & .004\\
                        .225 & .973 &  .041 \\
                        .009 & .040 &  .999 \\
                      \end{array}
                    \right).\label{|CKM|}
\end{eqnarray}
Up to three digits, the only deviation is the matrix element $|V_{td}|$, which is $.013$ rather than the measured  $.009$. With these choices of $\lambda$ and $\lambda^\prime$, we find $m_s/m_d\simeq 19$ and $m_b/m_s\simeq 35.5$, close to the measured values $17 \leq m_s/m_d \leq 22$ and $42 \leq m_b/m_s \leq 47$ \cite{CKM5}.

\emph{$CP$-violation correction in down-quark sector.---}

Continuing to assume that the down-quark mass matrix is Hermitian, we now include $CP$-violating phases:
\begin{equation}
m_{CP}(\lambda,\lambda^\prime)=\left(
\resizebox{.55\hsize}{!}{$
    \begin{array}{ccc}
      1 &  -2e^{i\Theta_{ds}} &  -\sqrt{2}e^{i\Theta_{db}} \\
      -2e^{-i\Theta_{ds}} & \lambda &  -\lambda\sqrt{2}e^{i\Theta_{sb}} \\
       -\sqrt{2}e^{-i\Theta_{db}} &  -\lambda\sqrt{2}e^{-i\Theta_{sb}} & \lambda^\prime \\
    \end{array}
$}
  \right).
\end{equation}
Because we have the freedom to redefine the phases of the right-handed and left-handed down-quark fields, the $CP$-violating angle and the down-quark masses depend on only the invariant linear combination of phases
\begin{eqnarray}
\Phi=\Theta_{db}-\Theta_{sb}+\Theta_{ds}+\pi \quad ({\rm mod}
~2\pi).\end{eqnarray}
The best fit to the CKM mixing angles is obtained by choosing  $\lambda=9.66$, $\lambda^\prime=341$ and $\Phi= 1.25$ rad(with the gauge choice $\Theta_{db}=-1.89$ rad, $\Theta_{ds}=\Theta_{sb}=\pi$), we find
\begin{eqnarray}
V_{CP}\simeq
                    \left(
                     \begin{array}{ccc}
                        .9743 & .2253  & .0042e^{-i(1.20)}\\
                        -.2252 & .9734 &  .0411 \\
                        .0088e^{-i(.48)} & -.0404 &  .9991 \\
                      \end{array}
                    \right),
\end{eqnarray}
which is consistent with Eq.(\ref{|CKM|}) and predicts the  $CP$-violating angle $\delta_{13}\simeq 1.2$ rad.

For these choices of  $\lambda$ and $\lambda^\prime$ we find the mass ratios $m_s/m_d\simeq 18$ and $m_b/m_s\simeq 36$; this value of $m_b/m_s\simeq 36$ is slightly smaller than the current experimental value \cite{CKM5}.

\emph{Up-quark sector correction.---}
We do somewhat better by including the correction to the CKM matrix coming from the diagonalization of the up-quark mass matrix. Because $m_c/m_u \simeq 500$ is nearly 30 times larger than $m_s/m_d\simeq 18$, this correction has little effect on the predicted value of $m_s/m_d$, but it does notable modify the prediction for $m_b/m_s$. We assume the up-quark mass matrix has the form  $m(\bar\lambda,\bar\lambda^\prime)=\bar V {\rm{diag}}(m_u,m_c,m_t) \bar V^\dagger$.
Using the experimental mass ratios  $m_c/m_u\simeq 554$ and $m_t/m_c\simeq136$ \cite{CKM5}, we fix $\bar\lambda=555$ and $\bar\lambda^\prime=75000$. By choosing $\Theta_{db}=-1.86$ rad, $\Theta_{ds}=\Theta_{sb}=\pi$, $\lambda=9.7$ and $\lambda^\prime=457$, the predicted CKM matrix is
\begin{eqnarray}
\resizebox{.9\hsize}{!}{$
\bar V^\dagger V_{CP}\simeq
                    \left(
                     \begin{array}{ccc}
                        .9743 & .2253  & .0012-.0030i\\
                        -.2252 & .9734 &  .0412 \\
                        .0081-.0031i & -.0404 &  .9992 \\
                      \end{array}
                    \right),
$}
\label{CKM-predict}
\end{eqnarray}
in four-digit agreement with the experimental data \cite{CKM5}:
\begin{eqnarray}
\resizebox{.9\hsize}{!}{$
V_{{\rm{CKM}}}\simeq
                    \left(
                     \begin{array}{ccc}
                        .9743 & .2253  & .0013-.0033i\\
                        -.2252 & .9734 &  .0412 \\
                        .0080-.0032i & -.0404 &  .9992 \\
                      \end{array}
                    \right).
$}
\end{eqnarray}
For this best-fit up-quark mass matrix we obtain $m_s/m_d \simeq 18$ and $m_b/m_s \simeq 47$, consistent with current experimental observations \cite{CKM5}. The sensitivity of the predicted CKM matrix to the phases in the up-quark mass matrix, which we have ignored so far, is discussed in the Supplementary Material.

\emph{LO approximation for PMNS matrix.---}
We assume that the small left-handed neutrino masses result from the see-saw mechanism \cite{seesaw1,seesaw2,seesaw3,seesaw4,seesaw5}, the complete $6\times 6$ mass matrix can be expressed as
\begin{eqnarray}
 M_{total}=\left(
   \begin{array}{cc}
     0 & m_D \\
     m_D^T & M \\
   \end{array}
 \right),\label{seesaw}
\end{eqnarray}
where $m_D$ is the 3 $\times$ 3 Dirac mass matrix and $M$ is the 3 $\times$ 3 Majorana mass matrix for the heavy right-handed neutrinos. (Majorana masses for the left-handed neutrinos vanish due to electroweak gauge symmetry.) For $m_D$ comparable to the electroweak symmetry breaking scale ($250~GeV$) and $M$ of order the GUT energy scale($10^{15}~GeV$), left-handed neutrino masses are of order $0.01 {-} 0.1~ eV$.

For a proper basis choice, $m_D$ is diagonal. Following \cite{Guneutrino}, we assume that $m_D={\rm{diag}}(m,m,m)$ and that $M$ in the LO approximation has the form (up to an overall scale)
\begin{eqnarray}
\resizebox{.9\hsize}{!}{$
M=\left(
    \begin{array}{ccc}
      1 &  -\sqrt{2} &  -\sqrt{2} \\
      -\sqrt{2} & 1 &  -2 \\
       -\sqrt{2} &  -2 & 1 \\
    \end{array}
  \right)=U \left(
                      \begin{array}{ccc}
                        \sqrt{5} & 0 & 0 \\
                        0 & -\sqrt{5} & 0 \\
                        0 & 0 & 3 \\
                      \end{array}
                    \right)U^T,
$}
\label{diag}
\end{eqnarray}
where
\begin{eqnarray}
U=\left(
                     \begin{array}{ccc}
                                             \sqrt{\frac{5+\sqrt{5}}{10}} &    \sqrt{\frac{5-\sqrt{5}}{10}} &0 \\
                     -\sqrt{\frac{5-\sqrt{5}}{20}} & \sqrt{\frac{5+\sqrt{5}}{20}} & -\frac{1}{\sqrt{2}} \\
                       -\sqrt{\frac{5-\sqrt{5}}{20}}& \sqrt{\frac{5+\sqrt{5}}{20}} & \frac{1}{\sqrt{2}} \\
                      \end{array}
                    \right).\label{mixing-neutrino}
\end{eqnarray}
Here $M$ matches Eq.(\ref{mass-matrix}) for $\lambda=\lambda'=1$, except that we have swapped the $12$ and $23$ entries by relabeling the generations.
The corresponding LO mixing angles are consistent with the so-called golden ratio (GR) pattern \cite{masssymmetry1,masssymmetry2}
\begin{eqnarray}
\theta_{23}=-45^\circ, \, \theta_{13}=0, \, \theta_{12}=\arctan\frac{\sqrt{5}-1}{2}\simeq 31.7^\circ,
\end{eqnarray}
which is reasonably close to current observations (the minus sign in $\theta_{23}$ can be eliminated with an appropriate gauge choice).

In the limit $m\ll M$, the PMNS mixing matrix for light neutrinos also takes the form Eq.(\ref{mixing-neutrino}) in the LO approximation.
However, for the light neutrinos the mass hierarchy is inverted, with $m_1/m_3=m_2/m_3=3/\sqrt{5}$. Using the measured difference of masses squared $|\Delta m_{23}^2|\simeq 2.4\times 10^{-3} ~eV^2$, we then obtain $m_1=m_2\simeq 0.074~eV$ and $m_3\simeq 0.055~eV$ in the LO approximation. We can also estimate the effective mass scale for neutrinoless double beta decay $m_{0\nu\beta\beta}\equiv|\sum_{i=1}^3 U_{ei}^2m_i|\simeq 0.033~eV$. The negative eigenvalue of Eq.(\ref{diag}), which is not a mere gauge choice, significantly suppresses this quantity.

\emph{Symmetries of mass matrix. ---}
The mass matrix Eq.(\ref{diag}) is invariant under the three symmetry operations
\begin{eqnarray}
P&=&\left(
    \begin{array}{ccc}
      1 & 0 & 0 \\
      0 & 0 & 1 \\
      0 & 1 & 0 \\
    \end{array}
  \right),\nonumber\\
S&=&\frac{1}{\sqrt{5}}\left(
    \begin{array}{ccc}
      1 & -\sqrt{2} & -\sqrt{2} \\
      -\sqrt{2} & -\frac{(\sqrt{5}+1)}{2} & \frac{(\sqrt{5}-1)}{2} \\
      -\sqrt{2} & \frac{(\sqrt{5}-1)}{2}  &-\frac{(\sqrt{5}+1)}{2}\\
    \end{array}
  \right),\nonumber\\
  R&=&\frac{1}{\sqrt{2}}\left(
    \begin{array}{ccc}
      0 & i & i \\
      -i& \frac{1}{\sqrt{2}}  & -\frac{1}{\sqrt{2}} \\
      -i & -\frac{1}{\sqrt{2}} & \frac{1}{\sqrt{2}} \\
    \end{array}
  \right),
\end{eqnarray}
which satisfy
\begin{eqnarray}
P^TMP=M, \quad S^TMS=M, \quad R^TMR=M,
\end{eqnarray}
and
\begin{eqnarray}\label{D4-relations}
P^2&=&1,\quad S^2=1,\quad R^2=1,\nonumber \\ PS&=&SP,\quad PR=RP,\quad SR=-PRS.
\end{eqnarray}
The relations Eq.(\ref{D4-relations}) define the group $D_4$ (symmetry of the square), where we regard $-P$ as a rotation by $\pi$, $R$ as a reflection about an axis through the square's diagonal, and $S$ as a reflection about an axis that bisects opposite sides of the square. The most general right-handed neutrino mass matrix with this $D_4$ symmetry is

\begin{eqnarray}
M_{{\rm{General}}}=\left(
    \begin{array}{ccc}
      1 &  -\sqrt{2} &  -\sqrt{2} \\
      -\sqrt{2} & \alpha &  -(1+\alpha) \\
       -\sqrt{2} &  -(1+\alpha) & \alpha \\
    \end{array}\right).
\end{eqnarray}
The corresponding mixing angles do not depend on $\alpha$, while the mass eigenvalues become
$\sqrt{5},-\sqrt{5},1+2\alpha$. The first and second generation neutrinos transform as a two-dimensional irreducible representation of $D_4$; hence the degenerate masses. Compatibility with the Ansatz Eq.(\ref{mass-matrix}), after swapping the first and third generations, requires $\alpha = 1$. The symmetries $P$ and $S$ were discussed in \cite{masssymmetry1,masssymmetry2}, but the complex transformation $R$ was introduced in  \cite{Guneutrino}.


\emph{$CP$-violation correction in neutrino sector. ---}
In the LO approximation, where $M$ is real, symmetries enforce the degeneracy $m_1 = m_2$. Now we relax the symmetry, allowing phases in $M$ which split the degeneracy and generate a nonzero $\theta_{13}$. Hence within our scheme the observed nonzero $\Delta m_{21}^2$ and $\theta_{13}$ already provide evidence for $CP$ violation in the lepton sector.

In general the symmetric Majorana mass matrix $M_{CP}$ for the right-handed neutrinos has six independent phases, but three can eliminated by gauge fixing. Thus we now assume that $M_{CP}$ has the form
\begin{eqnarray}
M_{CP}=\left(
    \begin{array}{ccc}
      1 &  -\sqrt{2} &  -\sqrt{2} \\
      -\sqrt{2} & e^{i\Theta} &  -2e^{i\Phi} \\
       -\sqrt{2}  &  -2e^{i\Phi} & e^{i\Theta^\prime} \\
    \end{array}\right).\label{RHnu-phase}
\end{eqnarray}
By diagonalizing the $6\times 6$ mass matrix Eq.(\ref{seesaw}), assuming Eq.(\ref{RHnu-phase}) and $m_D = {\rm diag}(m,m,m)$, we may derive the three mass eigenvalues $m_1,m_2$ and $m_3$ for the light left-handed neutrinos and the corresponding PMNS mixing matrix.

\begin{figure}[tbp]
\vskip -0.5cm
\begin{center}
\hspace*{-1cm}
\subfigure[]{\includegraphics[width=0.5\columnwidth]{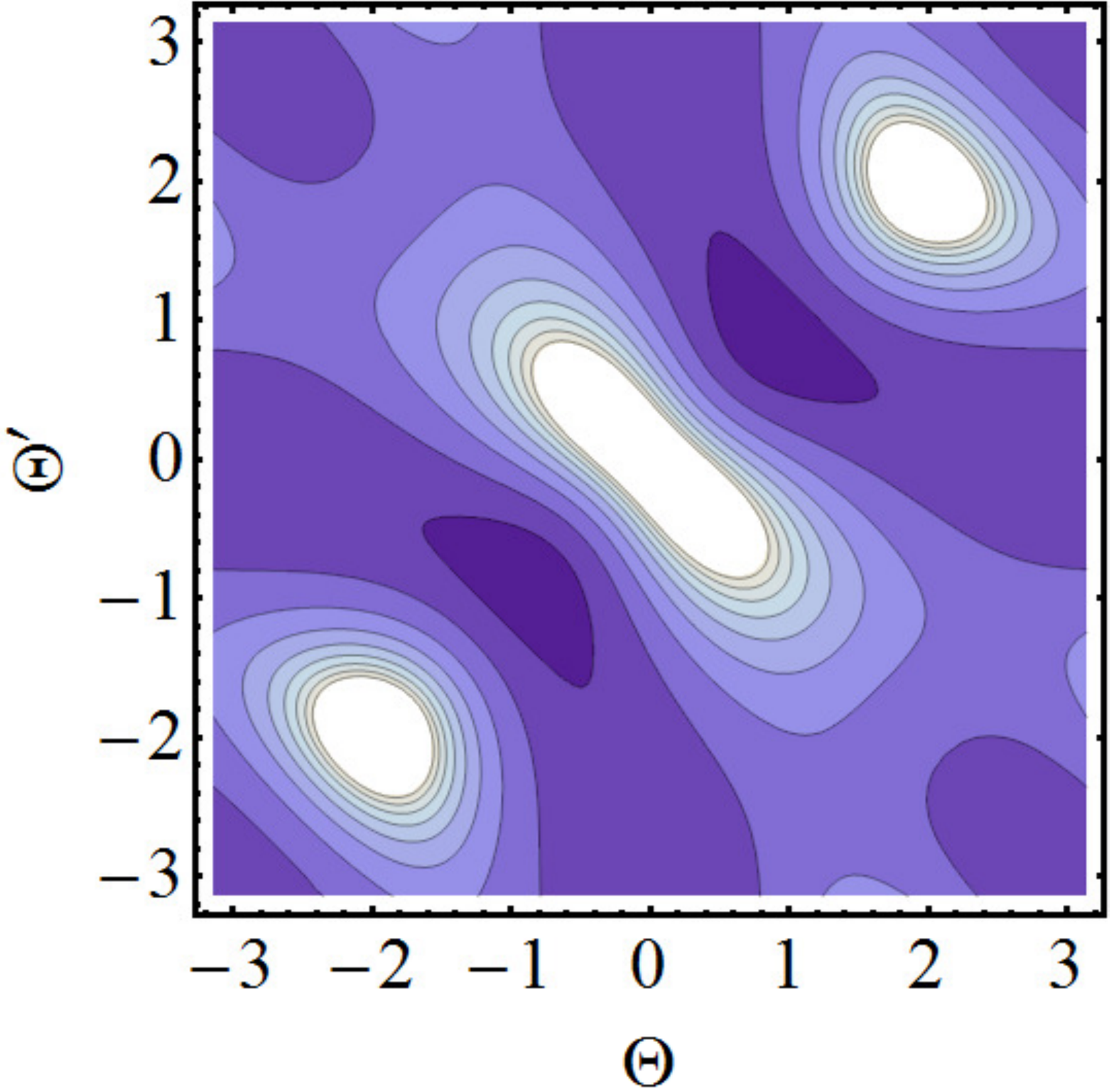}\label{ratio1}}
\hspace*{-0.8cm}
\subfigure[]{\includegraphics[width=0.65\columnwidth]{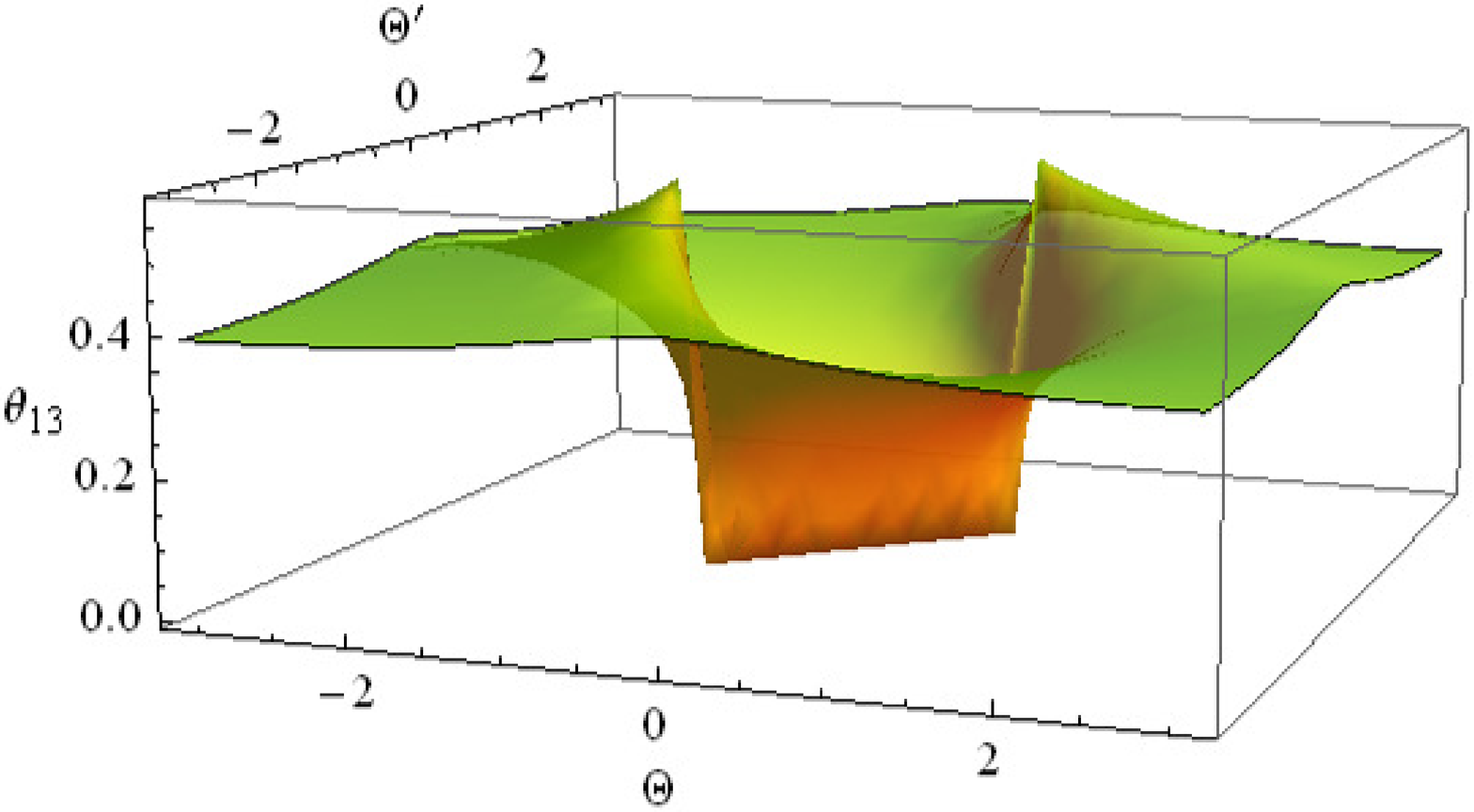}\label{angle1}}
\end{center}
\vskip -0.8cm
\caption{(Color online) (a) Contour plot for the mass splitting ratio $|\Delta m_{32}^2|/|\Delta m_{21}^2|$ as a function of $\Theta$ and $\Theta^\prime$ for $\Phi=0$. Only in white regions this ratio can be larger than 2(this ratio is infinite at origin due to $\Delta m_{21}^2=0$). (b) $\theta_{13}$ as a function of $\Theta$ and $\Theta^\prime$.\label{fig1}}
\end{figure}

\begin{figure}[tbp]
\vskip -0.5cm
\begin{center}
\hspace{-1cm}
\subfigure[]{\includegraphics[width=0.5\columnwidth]{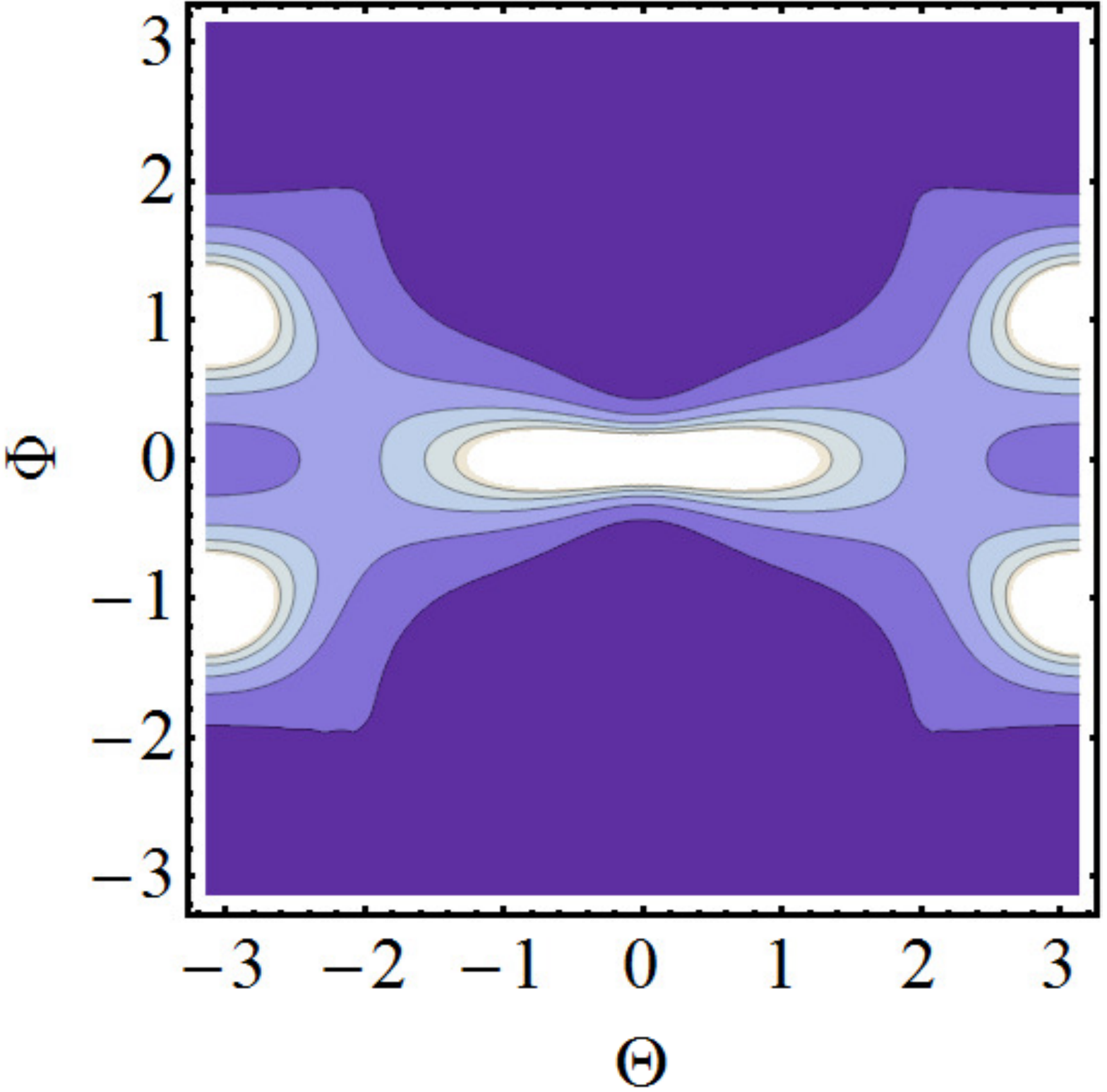}\label{ratio2}}
\hspace{-0.8cm}
\subfigure[]{\includegraphics[width=0.65\columnwidth]{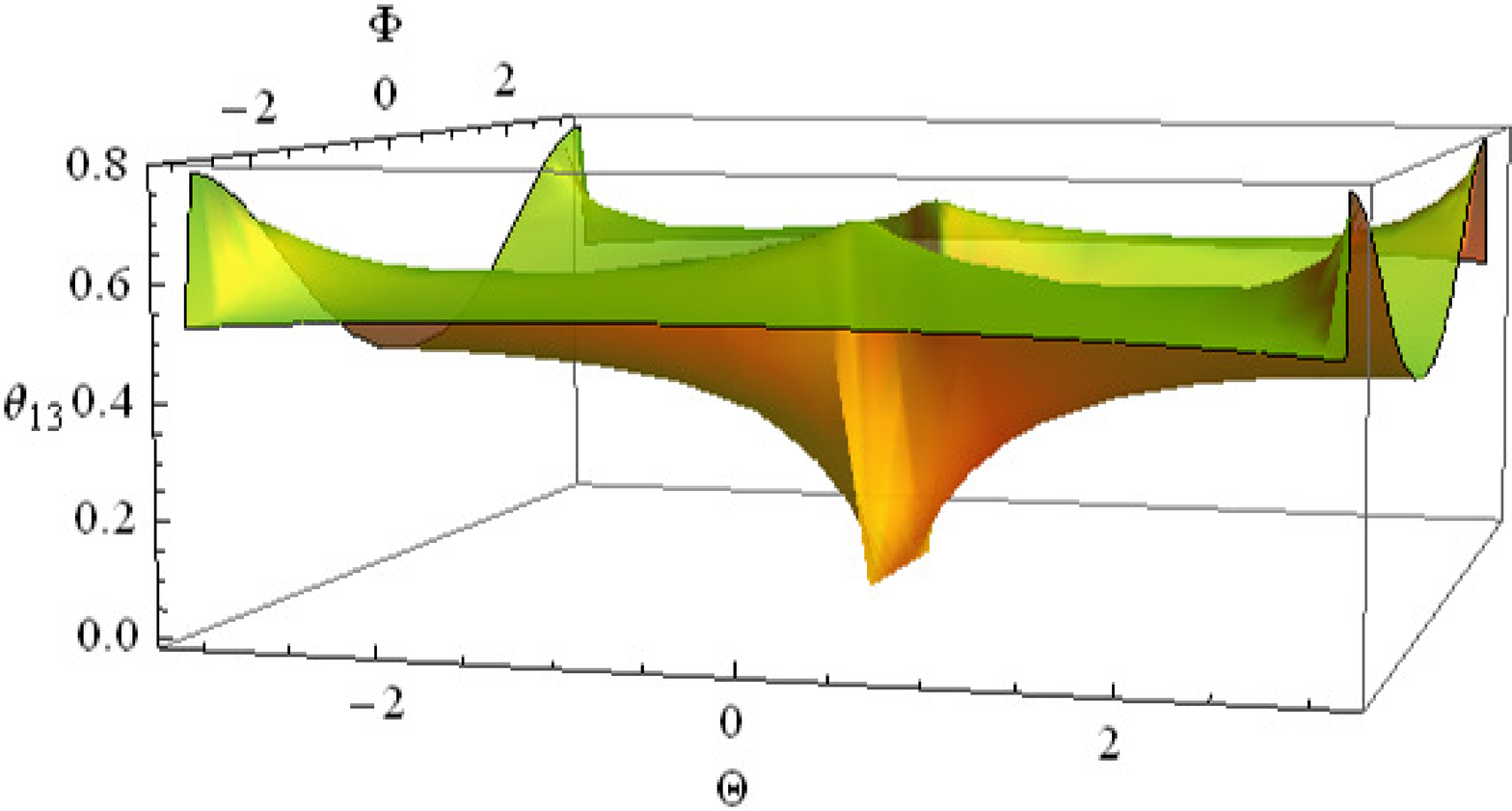}\label{angle2}}
\end{center}
\vskip -0.8cm
\caption{(Color online) (a) Contour plot for the mass splitting ratio $|\Delta m_{32}^2|/|\Delta m_{21}^2|$ as a function of $\Theta(=-\Theta^\prime)$ and $\Phi$. Only in white regions this ratio can be larger than 1(this ratio is infinite at origin due to $\Delta m_{21}^2=0$). (b) $\theta_{13}$ as a function of $\Theta(=-\Theta^\prime)$ and $\Phi$.\label{fig2}}
\end{figure}

We find that the large observed mass splitting ratio $|\Delta m_{32}^2|/|\Delta m_{21}^2|$($\simeq 30-33$) and the relatively small mixing angle $\theta_{13}$($\simeq 0.15$ rad are consistent with Eq.(\ref{RHnu-phase}) only for $\Theta\sim-\Theta^\prime$ and $\Phi\sim 0$. This constraint is illustrated in Fig. \ref{fig1} and Fig. \ref{fig2}, where we plot $|\Delta m_{32}^2|/|\Delta m_{21}^2|$ and $\theta_{13}$ as a function of $\Theta$, $\Theta'$ for $\Phi$ fixed, and as a function of $\Theta$ and $\Phi$ for $\Theta = -\Theta'$.

When we choose $\Theta = -\Theta^\prime$ and $\Phi= 0$, the mass matrix
\begin{eqnarray}
M_{SCP}=\left(
    \begin{array}{ccc}
      1 &  -\sqrt{2} &  -\sqrt{2} \\
      -\sqrt{2} & e^{i\Theta} &  -2 \\
       -\sqrt{2}  &  -2 & e^{-i\Theta} \\
    \end{array}\right)\label{MSCP}
\end{eqnarray}
has a residual $\mathbb{Z}_2$ anti-unitary symmetry generated by $\t P =PK$,
where $K$ is complex conjugation. It is important that this surviving $\mu$-$\tau$ $\mathbb{Z}_2$ symmetry is anti-unitary; if it were unitary instead, the CP violating phase in the mass matrix would not generate a nonzero $\theta_{13}$\cite{unitaryZ2}.
Assuming this unbroken symmetry in the neutrino sector, the angle $\Theta$ must be $\simeq \pm \pi/18$ to account for the measured value of $|\Delta m_{32}^2|/|\Delta m_{21}^2| \simeq 33$. With $\Theta$ fixed, the mixing angle is predicted to be $\theta_{13}\simeq 4.8^\circ$; this predicted value is smaller than the value measured by Daya Bay \cite{neutrinoexp9} by about three sigma, and deviates by more than three sigma from the best global fit to $\theta_{13}$ \cite{CKM5}.
With a proper gauge choice (see Supplementary Material for details), we find the $\mathbb{Z}_2$ anti-unitary symmetry leads to a maximal $CP$-violating angle $\delta=\pm \pi/2$. We also find $\alpha_1=0$, $\alpha_2=\pi$, which may be tested in future experiments.

We note that the agreement with the measured value of $\theta_{13}$ can be improved by relaxing the $D_4$ symmetry of the LO neutrino mass matrix, while maintaining the $\mathbb{Z}_2$ antiunitary symmetry, and that this $\mathbb{Z}_2$ symmetry alone leads to some robust predictions.   For example, if we choose the LO matrix $m(\lambda,\lambda^\prime)$ with $\lambda=1$ and $\lambda^\prime=0.96$, the $\mathbb{Z}_2$ anti-unitary symmetry is preserved. As for Eq.(\ref{MSCP}), we then choose $\Theta \simeq \pm \pi/9$ to fit $|\Delta m_{32}^2|/|\Delta m_{21}^2| \simeq 32$ and obtain $\theta_{13} \sim 9.2^\circ$, which is close to the measured value \cite{neutrinoexp9,CKM5}.
We emphasize that the $\mathbb{Z}_2$ anti-unitary symmetry alone enforces our predictions for
$\delta$, $\alpha_1$ and $\alpha_2$.

\emph{Charged-lepton sector correction.---}
We also parameterize the charged-lepton mass matrix as $m(\t \lambda,\t \lambda^\prime)=\t U{\rm{diag}}(m_e,m_\mu,m_\tau) \t U^\dagger$, and determine $\t \lambda=230$ and $\t \lambda^\prime=4000$ by fitting the mass ratios $m_\mu/m_e \simeq 206$ and $m_\tau/m_\mu \simeq 17$ \cite{CKM5}. The PMNS mixing matrix then becomes $\t U^\dagger U_{CP}$. This charged-lepton correction has only a negligible effect on $\theta_{12}$, $\theta_{13}$ and $\delta$, because the mass ratios are large, but $\theta_{23}$ is slightly modified to $\simeq 40.1^\circ$, which is reasonably close to the best global fit \cite{CKM5}.
Including phase factors in the charged-lepton mass matrix also has little effect on the other mixing angles or the $CP$-violating phase, but can change $\theta_{23}$, as we discuss in the Supplementary Material.

\emph{Conclusions and discussions.---}
We have proposed a unified form for quark and lepton mass matrices, which applies in a ``leading order'' approximation where $CP$-violating phases are ignored. Down-quark mass ratios are successfully predicted in our scheme using the measured CKM mixing angles as input. For neutrinos we predict a ``golden ratio'' mixing pattern in leading order, and  an inverted mass hierarchy with $m_1\simeq m_2 \simeq 74 ~meV$, $m_3\simeq 55~ meV$; for the neutrinoless double beta decay mass parameter we predict $m_{0\nu\beta\beta}\simeq 33~ meV$. We also predict that the $CP$-violating angle in the neutrino sector is close to the maximal value $\delta=\pm\pi/2$, and that the phases on the diagonal of the PMNS mixing matrix are $\alpha_1=0$, $\alpha_2=\pi$.

\emph{Acknowledgments.---} We thank Ryan Patterson for helpful discussions.
ZCG is supported by the Government of Canada through Industry Canada and by the Province of Ontario through the Ministry of
Research and Innovation.  JP is supported by NSF grant PHY-0803371 and DOE grant DE-FG03-92-ER40701. 
\bibliography{CKM}

\begin{thebibliography}{34}
\expandafter\ifx\csname natexlab\endcsname\relax\def\natexlab#1{#1}\fi
\expandafter\ifx\csname bibnamefont\endcsname\relax
  \def\bibnamefont#1{#1}\fi
\expandafter\ifx\csname bibfnamefont\endcsname\relax
  \def\bibfnamefont#1{#1}\fi
\expandafter\ifx\csname citenamefont\endcsname\relax
  \def\citenamefont#1{#1}\fi
\expandafter\ifx\csname url\endcsname\relax
  \def\url#1{\texttt{#1}}\fi
\expandafter\ifx\csname urlprefix\endcsname\relax\def\urlprefix{URL }\fi
\providecommand{\bibinfo}[2]{#2}
\providecommand{\eprint}[2][]{\url{#2}}

\bibitem[{\citenamefont{Cabibbo}(1963)}]{CKM1}
\bibinfo{author}{\bibfnamefont{N.}~\bibnamefont{Cabibbo}},
  \bibinfo{journal}{Phys. Rev. Lett.} \textbf{\bibinfo{volume}{10}},
  \bibinfo{pages}{351} (\bibinfo{year}{1963}).

\bibitem[{\citenamefont{Kobayashi and Maskawa}(1973)}]{CKM2}
\bibinfo{author}{\bibfnamefont{M.}~\bibnamefont{Kobayashi}} \bibnamefont{and}
  \bibinfo{author}{\bibfnamefont{T.}~\bibnamefont{Maskawa}},
  \bibinfo{journal}{Prog. Theor. Phys.} \textbf{\bibinfo{volume}{49}},
  \bibinfo{pages}{652} (\bibinfo{year}{1973}).

\bibitem[{\citenamefont{Charles et~al.}(2005)\citenamefont{Charles, Hocker,
  Lacker, Laplace, Diberder, Malcl¨¦s, Ocariz, Pivk, and Roos}}]{CKM3}
\bibinfo{author}{\bibfnamefont{J.}~\bibnamefont{Charles}},
  \bibinfo{author}{\bibfnamefont{A.}~\bibnamefont{Hocker}},
  \bibinfo{author}{\bibfnamefont{H.}~\bibnamefont{Lacker}},
  \bibinfo{author}{\bibfnamefont{S.}~\bibnamefont{Laplace}},
  \bibinfo{author}{\bibfnamefont{F.~R.~L.} \bibnamefont{Diberder}},
  \bibinfo{author}{\bibfnamefont{J.}~\bibnamefont{Malcl¨¦s}},
  \bibinfo{author}{\bibfnamefont{J.}~\bibnamefont{Ocariz}},
  \bibinfo{author}{\bibfnamefont{M.}~\bibnamefont{Pivk}}, \bibnamefont{and}
  \bibinfo{author}{\bibfnamefont{L.}~\bibnamefont{Roos}}, \bibinfo{journal}{The
  European Physical Journal C - Particles and Fields}
  \textbf{\bibinfo{volume}{41}}, \bibinfo{pages}{1} (\bibinfo{year}{2005}).

\bibitem[{\citenamefont{Beringer et~al.}(2012)}]{CKM5}
\bibinfo{author}{\bibfnamefont{J.}~\bibnamefont{Beringer}}
  \bibnamefont{et~al.}, \bibinfo{journal}{(Particle Data Group)Phys. Rev. D}
  \textbf{\bibinfo{volume}{86}}, \bibinfo{pages}{010001}
  (\bibinfo{year}{2012}).

\bibitem[{\citenamefont{Fukuda et~al.}(1998)}]{neutrinoexp1}
\bibinfo{author}{\bibfnamefont{Y.}~\bibnamefont{Fukuda}} \bibnamefont{et~al.},
  \bibinfo{journal}{Phys. Rev. Lett} \textbf{\bibinfo{volume}{81}},
  \bibinfo{pages}{1562} (\bibinfo{year}{1998}),
  \bibinfo{note}{[Super-Kamiokande Collaboration]}.

\bibitem[{\citenamefont{Ahmad et~al.}(2002)}]{neutrinoexp2}
\bibinfo{author}{\bibfnamefont{Q.~R.} \bibnamefont{Ahmad}}
  \bibnamefont{et~al.}, \bibinfo{journal}{Phys. Rev. Lett}
  \textbf{\bibinfo{volume}{89}}, \bibinfo{pages}{011301}
  (\bibinfo{year}{2002}), \bibinfo{note}{[SNO Collaboration]}.

\bibitem[{\citenamefont{Ahmed et~al.}(2004)}]{neutrinoexp3}
\bibinfo{author}{\bibfnamefont{S.~N.} \bibnamefont{Ahmed}}
  \bibnamefont{et~al.}, \bibinfo{journal}{Phys. Rev. Lett}
  \textbf{\bibinfo{volume}{92}}, \bibinfo{pages}{181301}
  (\bibinfo{year}{2004}), \bibinfo{note}{[SNO Collaboration]}.

\bibitem[{\citenamefont{Eguchi et~al.}(2003)}]{neutrinoexp4}
\bibinfo{author}{\bibfnamefont{K.}~\bibnamefont{Eguchi}} \bibnamefont{et~al.},
  \bibinfo{journal}{Phys. Rev. Lett} \textbf{\bibinfo{volume}{90}},
  \bibinfo{pages}{021802} (\bibinfo{year}{2003}), \bibinfo{note}{[KamLAND
  Collaboration]}.

\bibitem[{\citenamefont{Ahn et~al.}(2006)}]{neutrinoexp5}
\bibinfo{author}{\bibfnamefont{M.~H.} \bibnamefont{Ahn}} \bibnamefont{et~al.},
  \bibinfo{journal}{Phys. Rev. D} \textbf{\bibinfo{volume}{74}},
  \bibinfo{pages}{072003} (\bibinfo{year}{2006}), \bibinfo{note}{[K2K
  Collaboration]}.

\bibitem[{\citenamefont{Abe et~al.}(2011)}]{neutrinoexp6}
\bibinfo{author}{\bibfnamefont{K.}~\bibnamefont{Abe}} \bibnamefont{et~al.},
  \bibinfo{journal}{Phys. Rev. Lett} \textbf{\bibinfo{volume}{107}},
  \bibinfo{pages}{041801} (\bibinfo{year}{2011}), \bibinfo{note}{[T2K
  Collaboration]}.

\bibitem[{\citenamefont{Adamson et~al.}(2011)}]{neutrinoexp7}
\bibinfo{author}{\bibfnamefont{P.}~\bibnamefont{Adamson}} \bibnamefont{et~al.},
  \bibinfo{journal}{Phys. Rev. Lett} \textbf{\bibinfo{volume}{107}},
  \bibinfo{pages}{181802} (\bibinfo{year}{2011}), \bibinfo{note}{[MINOS
  Collaboration]}.

\bibitem[{\citenamefont{Abe et~al.}(2012{\natexlab{a}})}]{neutrinoexp8}
\bibinfo{author}{\bibfnamefont{Y.}~\bibnamefont{Abe}} \bibnamefont{et~al.},
  \bibinfo{journal}{Phys. Rev. Lett} \textbf{\bibinfo{volume}{108}},
  \bibinfo{pages}{131801} (\bibinfo{year}{2012}{\natexlab{a}}),
  \bibinfo{note}{[DOUBLE-CHOOZ Collaboration]}.

\bibitem[{\citenamefont{An et~al.}(2012)}]{neutrinoexp9}
\bibinfo{author}{\bibfnamefont{F.~P.} \bibnamefont{An}} \bibnamefont{et~al.},
  \bibinfo{journal}{Phys. Rev. Lett} \textbf{\bibinfo{volume}{108}},
  \bibinfo{pages}{171803} (\bibinfo{year}{2012}), \bibinfo{note}{[DAYA-BAY
  Collaboration]}.

\bibitem[{\citenamefont{Ahn et~al.}(2012)}]{neutrinoexp10}
\bibinfo{author}{\bibfnamefont{J.~K.} \bibnamefont{Ahn}} \bibnamefont{et~al.},
  \bibinfo{journal}{Phys. Rev. Lett} \textbf{\bibinfo{volume}{108}},
  \bibinfo{pages}{191802} (\bibinfo{year}{2012}), \bibinfo{note}{[RENO
  Collaboration]}.

\bibitem[{\citenamefont{Abe et~al.}(2012{\natexlab{b}})}]{neutrinoexp11}
\bibinfo{author}{\bibfnamefont{Y.}~\bibnamefont{Abe}} \bibnamefont{et~al.},
  \bibinfo{journal}{Phys. Rev. D} \textbf{\bibinfo{volume}{86}},
  \bibinfo{pages}{052008} (\bibinfo{year}{2012}{\natexlab{b}}),
  \bibinfo{note}{[Double Chooz Collaboration]}.

\bibitem[{\citenamefont{Abe et~al.}(2014)}]{neutrinoexp12}
\bibinfo{author}{\bibfnamefont{K.}~\bibnamefont{Abe}} \bibnamefont{et~al.},
  \bibinfo{journal}{Phys. Rev. Lett} \textbf{\bibinfo{volume}{112}},
  \bibinfo{pages}{061802} (\bibinfo{year}{2014}), \bibinfo{note}{(T2K
  Collaboration)}.

\bibitem[{\citenamefont{Raidal}(2004)}]{complementarity1}
\bibinfo{author}{\bibfnamefont{M.}~\bibnamefont{Raidal}},
  \bibinfo{journal}{Phys. Rev. Lett.} \textbf{\bibinfo{volume}{93}},
  \bibinfo{pages}{161801} (\bibinfo{year}{2004}).

\bibitem[{\citenamefont{Minakata1 and Smirnov1}(2004)}]{complementarity2}
\bibinfo{author}{\bibfnamefont{H.}~\bibnamefont{Minakata1}} \bibnamefont{and}
  \bibinfo{author}{\bibfnamefont{A.~Y.} \bibnamefont{Smirnov1}},
  \bibinfo{journal}{Phys. Rev. D} \textbf{\bibinfo{volume}{70}},
  \bibinfo{pages}{073009} (\bibinfo{year}{2004}).

\bibitem[{\citenamefont{Babua et~al.}(2003)\citenamefont{Babua, Mab, and
  Vallec}}]{symmetry1}
\bibinfo{author}{\bibfnamefont{K.}~\bibnamefont{Babua}},
  \bibinfo{author}{\bibfnamefont{E.}~\bibnamefont{Mab}}, \bibnamefont{and}
  \bibinfo{author}{\bibfnamefont{J.}~\bibnamefont{Vallec}},
  \bibinfo{journal}{Physics Letters B} \textbf{\bibinfo{volume}{552}},
  \bibinfo{pages}{207} (\bibinfo{year}{2003}).

\bibitem[{\citenamefont{He et~al.}(2006)\citenamefont{He, Keum, and
  Volkas}}]{symmetry2}
\bibinfo{author}{\bibfnamefont{X.-G.} \bibnamefont{He}},
  \bibinfo{author}{\bibfnamefont{Y.-Y.} \bibnamefont{Keum}}, \bibnamefont{and}
  \bibinfo{author}{\bibfnamefont{R.~R.} \bibnamefont{Volkas}},
  \bibinfo{journal}{JHEP} \textbf{\bibinfo{volume}{04}}, \bibinfo{pages}{039}
  (\bibinfo{year}{2006}).

\bibitem[{\citenamefont{Feruglio et~al.}(2007)\citenamefont{Feruglio, Hagedorn,
  Lin, and Merlo}}]{symmetry3}
\bibinfo{author}{\bibfnamefont{F.}~\bibnamefont{Feruglio}},
  \bibinfo{author}{\bibfnamefont{C.}~\bibnamefont{Hagedorn}},
  \bibinfo{author}{\bibfnamefont{Y.}~\bibnamefont{Lin}}, \bibnamefont{and}
  \bibinfo{author}{\bibfnamefont{L.}~\bibnamefont{Merlo}},
  \bibinfo{journal}{Nucl.Phys.B} \textbf{\bibinfo{volume}{775}},
  \bibinfo{pages}{120} (\bibinfo{year}{2007}).

\bibitem[{\citenamefont{Altarelli and Feruglio}(2010)}]{masssymmetry3}
\bibinfo{author}{\bibfnamefont{G.}~\bibnamefont{Altarelli}} \bibnamefont{and}
  \bibinfo{author}{\bibfnamefont{F.}~\bibnamefont{Feruglio}},
  \bibinfo{journal}{Rev. Mod. Phys.} \textbf{\bibinfo{volume}{82}},
  \bibinfo{pages}{2701} (\bibinfo{year}{2010}).

\bibitem[{\citenamefont{King and Luhn}(2013)}]{masssymmetry4}
\bibinfo{author}{\bibfnamefont{S.~F.} \bibnamefont{King}} \bibnamefont{and}
  \bibinfo{author}{\bibfnamefont{C.}~\bibnamefont{Luhn}},
  \bibinfo{journal}{arXiv:1301.1340}  (\bibinfo{year}{2013}).

\bibitem[{\citenamefont{Gu}(2013)}]{Guneutrino}
\bibinfo{author}{\bibfnamefont{Z.-C.} \bibnamefont{Gu}},
  \bibinfo{journal}{arXiv:1308.2488}  (\bibinfo{year}{2013}).

\bibitem[{\citenamefont{Gibbons
  et~al.}(2009{\natexlab{a}})\citenamefont{Gibbons, Gielen, Pope, and
  Turok}}]{NeilCKM1}
\bibinfo{author}{\bibfnamefont{G.~W.} \bibnamefont{Gibbons}},
  \bibinfo{author}{\bibfnamefont{S.}~\bibnamefont{Gielen}},
  \bibinfo{author}{\bibfnamefont{C.~N.} \bibnamefont{Pope}}, \bibnamefont{and}
  \bibinfo{author}{\bibfnamefont{N.}~\bibnamefont{Turok}},
  \bibinfo{journal}{Phys. Rev. Lett.} \textbf{\bibinfo{volume}{102}},
  \bibinfo{pages}{121802} (\bibinfo{year}{2009}{\natexlab{a}}).

\bibitem[{\citenamefont{Gibbons
  et~al.}(2009{\natexlab{b}})\citenamefont{Gibbons, Gielen, Pope, and
  Turok}}]{NeilCKM2}
\bibinfo{author}{\bibfnamefont{G.~W.} \bibnamefont{Gibbons}},
  \bibinfo{author}{\bibfnamefont{S.}~\bibnamefont{Gielen}},
  \bibinfo{author}{\bibfnamefont{C.~N.} \bibnamefont{Pope}}, \bibnamefont{and}
  \bibinfo{author}{\bibfnamefont{N.}~\bibnamefont{Turok}},
  \bibinfo{journal}{Phys. Rev. D} \textbf{\bibinfo{volume}{79}},
  \bibinfo{pages}{013009} (\bibinfo{year}{2009}{\natexlab{b}}).

\bibitem[{\citenamefont{Gell-Mann et~al.}(1979)\citenamefont{Gell-Mann, Ramond,
  and Slansky}}]{seesaw1}
\bibinfo{author}{\bibfnamefont{M.}~\bibnamefont{Gell-Mann}},
  \bibinfo{author}{\bibfnamefont{P.}~\bibnamefont{Ramond}}, \bibnamefont{and}
  \bibinfo{author}{\bibfnamefont{R.}~\bibnamefont{Slansky}},
  \bibinfo{journal}{in Sanibel Talk, CALT-68-709, Feb 1979, and in
  Supergravity, North Holland, Amsterdam}  (\bibinfo{year}{1979}).

\bibitem[{\citenamefont{Yanagida}(1979)}]{seesaw2}
\bibinfo{author}{\bibfnamefont{T.}~\bibnamefont{Yanagida}},
  \bibinfo{journal}{in Proc. of the Workshop on Unified Theory and Baryon
  Number of the Universe, KEK, Japan}  (\bibinfo{year}{1979}).

\bibitem[{\citenamefont{Wienberg}(1979)}]{seesaw3}
\bibinfo{author}{\bibfnamefont{S.}~\bibnamefont{Wienberg}},
  \bibinfo{journal}{Phys. Rev. Lett.} \textbf{\bibinfo{volume}{43}},
  \bibinfo{pages}{1566} (\bibinfo{year}{1979}).

\bibitem[{\citenamefont{Mohapatra and Senjanovic}(1980)}]{seesaw4}
\bibinfo{author}{\bibfnamefont{R.~N.} \bibnamefont{Mohapatra}}
  \bibnamefont{and}
  \bibinfo{author}{\bibfnamefont{G.}~\bibnamefont{Senjanovic}},
  \bibinfo{journal}{Phys. Rev. Lett.} \textbf{\bibinfo{volume}{44}},
  \bibinfo{pages}{912} (\bibinfo{year}{1980}).

\bibitem[{\citenamefont{Schechter and Valle}(1980)}]{seesaw5}
\bibinfo{author}{\bibfnamefont{J.}~\bibnamefont{Schechter}} \bibnamefont{and}
  \bibinfo{author}{\bibfnamefont{J.~W.~F.} \bibnamefont{Valle}},
  \bibinfo{journal}{Phys. Rev. D} \textbf{\bibinfo{volume}{22}},
  \bibinfo{pages}{2227} (\bibinfo{year}{1980}).

\bibitem[{\citenamefont{Kajiyama et~al.}(2007)\citenamefont{Kajiyama, Raidal,
  and Strumia}}]{masssymmetry1}
\bibinfo{author}{\bibfnamefont{Y.}~\bibnamefont{Kajiyama}},
  \bibinfo{author}{\bibfnamefont{M.}~\bibnamefont{Raidal}}, \bibnamefont{and}
  \bibinfo{author}{\bibfnamefont{A.}~\bibnamefont{Strumia}},
  \bibinfo{journal}{Phys. Rev. D} \textbf{\bibinfo{volume}{82}},
  \bibinfo{pages}{117301} (\bibinfo{year}{2007}).

\bibitem[{\citenamefont{Feruglio and Paris}(2011)}]{masssymmetry2}
\bibinfo{author}{\bibfnamefont{F.}~\bibnamefont{Feruglio}} \bibnamefont{and}
  \bibinfo{author}{\bibfnamefont{A.}~\bibnamefont{Paris}},
  \bibinfo{journal}{arXiv:1101.0393}  (\bibinfo{year}{2011}).

\bibitem[{\citenamefont{Dicus et~al.}(2011)\citenamefont{Dicus, Ge, and
  Repko}}]{unitaryZ2}
\bibinfo{author}{\bibfnamefont{D.~A.} \bibnamefont{Dicus}},
  \bibinfo{author}{\bibfnamefont{S.-F.} \bibnamefont{Ge}}, \bibnamefont{and}
  \bibinfo{author}{\bibfnamefont{W.~W.} \bibnamefont{Repko}},
  \bibinfo{journal}{Phys. Rev. D} \textbf{\bibinfo{volume}{83}},
  \bibinfo{pages}{093007} (\bibinfo{year}{2011}).

\end{thebibliography}

\begin{widetext}
\section{Supplementary material}
\subsection{$CP$-violating phases in the up-quark sector}
When $CP$-violating phases are included, our Ansatz for the up-quark mass matrix takes the form
\begin{eqnarray}
m_{CP}(\bar\lambda,\bar\lambda^\prime)=\left(
    \begin{array}{ccc}
      1 &  -2e^{i\Theta_{uc}} &  -\sqrt{2}e^{i\Theta_{ut}} \\
      -2e^{-i\Theta_{uc}} & \bar\lambda &  -\bar\lambda\sqrt{2}e^{i\Theta_{ct}} \\
       -\sqrt{2}e^{-i\Theta_{ut}} &  -\bar\lambda\sqrt{2}e^{-i\Theta_{ct}} & \bar\lambda^\prime \\
    \end{array}
  \right).
\end{eqnarray}
By choosing $\bar\lambda=555$ and $\bar\lambda^\prime=75000$, we fit the experimentally observed mass ratios $m_c/m_u\simeq 554$ and $m_t/m_c\simeq136$ \cite{CKM5}.

Because the mass ratios $m_c/m_u$ and $m_c/m_u$ are quite large, our predicted CKM mixing matrix $\bar V^\dagger V_{CP}$ is nearly independent of the phases $\Theta_{uc}$ and $\Theta_{ut}$, but somewhat more sensitive to $\Theta_{ct}$ (because $m_t/m_c\simeq 136$ is not so much larger than $m_b/m_s \simeq 47$).

For $\Theta_{ct} = 0$, we can fit the measured CKM matrix well by choosing $\lambda= 9.7$ and $\lambda'=458$ as in Eq.(\ref{CKM-predict}), finding for these values the down-quark mass ratios $m_s/m_d \simeq 18$ and $m_b/m_s \simeq 47$, also in good agreement with experiment. To illustrate the sensitivity of the results to the value of $\Theta_{ct}$, consider $\Theta_{ct}=\pi$; then by choosing $\lambda=9.97$ and $\lambda^\prime=281$ (and choosing the gauge $\Theta_{db}=-1.84$ rad, $\Theta_{ds}=\Theta_{sb}=\pi$), we obtain the best fit to the CKM matrix:
\begin{eqnarray}
\bar V^\dagger V_{CP}\simeq
                    \left(
                     \begin{array}{ccc}
                        .9743 & .2253  & .0019-.0049i\\
                        -.2252 & .9734 &  .0412 \\
                        .0075-.0050i & -.0405 &  .9991 \\
                      \end{array}
                    \right).
\end{eqnarray}
This choice of $\lambda$ and $\lambda^\prime$ implies $m_s/m_d \simeq 18$ and $m_b/m_s \simeq 29$.

When $\Theta_{ct}=\pi/2$, we choose $\lambda=9.83$ and $\lambda^\prime=348$ (and the gauge choice $\Theta_{db}=-1.84$, $\Theta_{ds}=\Theta_{sb}=\pi$), to obtain the best fit to the CKM matrix:
\begin{eqnarray}
\bar V^\dagger V_{CP}\simeq
                    \left(
                     \begin{array}{ccc}
                        .9743 & .2253  & .0015-.0039i\\
                        -.2252 & .9734 &  .0410 \\
                        .0077-.0064i & -.0403 &  .9992 \\
                      \end{array}
                    \right).
\end{eqnarray}
This choice for $\lambda$ and $\lambda^\prime$ implies $m_s/m_d \simeq 18$ and $m_b/m_s \simeq 36$.

As $\Theta_{ct}$ varies continuously from $0$ to $\pm \pi$, we find that the mass ratio $m_s/m_d \simeq 18$ is nearly constant,  while $m_b/m_s$ ranges from $47$ to $29$. Thus the experimentally observed $m_b/m_s$ ($42\leq m_b/m_s\leq 47$) favors small $\Theta_{ct}$, and we also find a notably better fit to $V_{td}$ and $V_{ub}$ for small $\Theta_{ct}$.

\subsection{Gauge choice for the phases in the neutrino sector}

The standard decomposition of the PMNS mixing matrix is
\begin{eqnarray}
U=
\left(
                     \begin{array}{ccc}
                        U_{e1} & U_{e2}  & U_{e3}\\
                        U_{\mu 1} & U_{\mu 2}  & U_{\mu 3} \\
                        U_{\tau 1} & U_{\tau 2}  & U_{\tau 3} \\
                      \end{array}
                    \right)
=
                    \left(
                     \begin{array}{ccc}
                        1 & 0  & 0\\
                        0 & c_{23} &  s_{23} \\
                        0 & -s_{23} &  c_{23} \\
                      \end{array}
		\right)
                    \left(
                     \begin{array}{ccc}
                        c_{13} & 0  & s_{13}e^{-i\delta}\\
                        0 & 1 &  0 \\
                        -s_{13}e^{i\delta}& 0 &  c_{13} \\
                      \end{array}
                    \right)
                    \left(
                     \begin{array}{ccc}
                        c_{12} & s_{12}  & 0\\
                        -s_{12} & c_{12} &  0 \\
                        0& 0 &  1 \\
                      \end{array}
                    \right)
                    \left(
                     \begin{array}{ccc}
                        e^{i\alpha_1/2}& 0  & 0\\
                        0 & e^{i\alpha_2/2} &  0 \\
                        0& 0 &  1 \\
                      \end{array}
                    \right).\label{PMNS-convention}
\end{eqnarray}
With this convention, the $CP$-violating phase $\delta$ is the phase of the $U_{e3}$ matrix element.

To enforce the gauge condition Eq.(\ref{PMNS-convention}), we rotate the phases in the second and third generation by $e^{i\phi}$ and $e^{-i\phi}$ respectively, thus gauge transforming Eq.(\ref{MSCP}) to
\begin{eqnarray}
 M_{SCP}=\left(
    \begin{array}{ccc}
      1 &  -\sqrt{2}e^{i\phi} &  -\sqrt{2}e^{-i\phi} \\
      -\sqrt{2}e^{i\phi} & e^{i(2\phi+\Theta)} &  -2 \\
       -\sqrt{2}e^{-i\phi}  &  -2 & e^{-i(2\phi+\Theta)} \\
    \end{array}\right).\label{MSCP-gauge}
\end{eqnarray}
As discussed in the main text, we can fit the experimentally measured ratio of mass squared differences $|\Delta m_{32}^2|/|\Delta m_{21}^2|\simeq 33$  by choosing $\Theta=\mp\pi/18 \simeq \mp0.174$ in $M_{SCP}$ (recall that $m_i\sim 1/M_i$ where $M_{1,2,3}$ are the eigenvalues of $M_{SCP}$). By choosing $\phi=\pm0.171$, we can diagonalize $M_{SCP}$ as:
\begin{eqnarray}
 M_{SCP}= U_{CP} \left(
                      \begin{array}{ccc}
                        M_1 & 0 & 0 \\
                        0 & M_2 & 0 \\
                        0 & 0 & M_3 \\
                      \end{array}
                    \right)U^T_{CP}, \label{diagCP}
\end{eqnarray}
where
\begin{eqnarray}
U_{CP}&\simeq&
                    \left(
                     \begin{array}{ccc}
                        .851 & .518 i & \mp.084i\\
                        -.368\mp0.051i & \pm.031+.604i & .705 \\
                        -.368\pm.051i & \mp.031+.604i & -.705 \\
                      \end{array}
                    \right)\nonumber\\&=&
                     \left(
                     \begin{array}{ccc}
                        .851 & .518  & \mp.084i\\
                        -.368\mp.051i & .604\mp.031i & .705 \\
                        -.368\pm.051i & .604\pm.031i & -.705 \\
                      \end{array}
                    \right)                     \left(
                     \begin{array}{ccc}
                        1 & 0 &0\\
                        0 & i & 0\\
                        0 & 0 & 1\\
                      \end{array}
                    \right)\label{mixingCP}
\end{eqnarray}
Thus we conclude that $\delta=\pm \pi/2$, $\alpha_1=0$, and $\alpha_2=\pi$.

The right-handed neutrino mass matrix $M_{SCP}$ also implies $m_1\simeq m_2\simeq \frac{3}{\sqrt{5}}m_3$; using the observed $|\Delta m_{32}^2| \simeq 2.4\times 10^{-3} ~eV^2$, we obtain $m_1\simeq m_2\simeq 0.074~eV$ and $m_3\simeq 0.055~eV$. The mass parameter for neutrinoless double beta decay is $m_{0\nu\beta\beta}\equiv |\sum_i U_{ei}^2 m_i|=|(0.851^2-0.518^2)\times 0.074~eV- 0.084^2\times 0.055~eV|\simeq 0.033~eV$.

\subsection{The effect of phases in the charged-lepton mass matrix}
As for the down-quark and up-quark sectors, we may relax our Ansatz for the charged-lepton mass matrix by including phases and maintaining Hermiticity, obtaining
\begin{eqnarray}
m_{CP}(\t\lambda,\t\lambda^\prime)=\left(
    \begin{array}{ccc}
      1 &  -2e^{i\Theta_{e\mu}} &  -\sqrt{2}e^{i\Theta_{e\tau}} \\
      -2e^{-i\Theta_{e\mu}} & \t\lambda &  -\t\lambda\sqrt{2}e^{i\Theta_{\mu\tau}} \\
       -\sqrt{2}e^{-i\Theta_{e\tau}} &  -\t\lambda\sqrt{2}e^{-i\Theta_{\mu\tau}} & \t\lambda^\prime \\
    \end{array}
  \right).
\end{eqnarray}
We choose $\t \lambda=230$ and $\t \lambda^\prime=4000$ to obtain the best fit to the experimental observed mass ratios $m_\mu/m_e \simeq 206$ and $m_\tau/m_\mu \simeq 17$\cite{CKM5}.

Because the mass ratios $m_\mu/m_e$ and $m_\tau/m_e$ are quite large, our predicted PMNS mixing matrix $\t U^\dagger U_{CP}$ is nearly independent of the phases $\Theta_{e\mu}$ and $\Theta_{e\tau}$; however, $\theta_{23}$ depends significantly on $\Theta_{\mu\tau}$ because $m_\tau/m_\mu\simeq 17$ is not so large.

We find that as $\Theta_{\mu\tau}$ ranges from 0 to $\pi$, the mixing angle $\theta_{23}$ varies from $40.1^\circ$ to $50.9^\circ$, while the other mixing angles and the phases in the PMNS matrix hardly vary at all. The dependence on the neutrinoless double beta decay mass scale $m_{0\nu\beta\beta}$ on the phases in the charged-lepton mass matrix is also negligible.

\end{widetext}
\end{document}